\documentclass[twocolumn,aps,prb,showpacs,amsfonts,amssymb,amsmath]{revtex4-1}
\usepackage{mathrsfs}
\usepackage{amsmath}
\usepackage{CJK}
\usepackage{graphicx}

\begin{document}
\title{Theory of orbital magnetization in disordered systems}
\author{Guobao Zhu$^{1}$}
\author{Shengyuan A. Yang$^{2,1}$}
\author{Cheng Fang$^{3,1}$}
\author{W. M. Liu$^{1}$}
\thanks{To whom correspondence should be addressed.\\  \texttt{ygyao@bit.edu.cn,ygyao@aphy.iphy.ac.cn}\\
        To whom correspondence should be addressed.\\  \texttt{wmliu@iphy.ac.cn}}
\author{Yugui Yao$^{4,1}$}
\thanks{To whom correspondence should be addressed.\\  \texttt{ygyao@bit.edu.cn,ygyao@aphy.iphy.ac.cn}\\
        To whom correspondence should be addressed.\\  \texttt{wmliu@iphy.ac.cn}}
\address{$^1$Beijing National Laboratory for Condensed Matter Physics, Institute of Physics, Chinese Academy of Sciences, Beijing
100190, China}
\address{$^2$Department of Physics, The University of Texas, Austin, 78712, USA}
\address{$^3$School of Science, East China Institute of Technology, Fuzhou 344000, China}
\address{$^4$School of Physics, Beijing Institute of Technology, Beijing 100081, China}

\date{\today}

\begin{abstract}
We present a general formula of the orbital magnetization of disordered systems based on the Keldysh Green's
function theory in the gauge-covariant Wigner space. In our approach, the gauge invariance of physical quantities is
ensured from the very beginning, and the vertex corrections are easily included.
Our formula applies not only for insulators but also for metallic systems where the quasiparicle behavior is
usually strongly modified by the disorder scattering.
In the absence of disorders, our formula recovers the previous results obtained from the semiclassical theory
and the perturbation theory. As an application, we calculate the orbital magnetization of
a weakly disordered two-dimensional electron gas with Rashba spin-orbit coupling.
We find that for the short range disorder scattering, its major effect is to the shifting of the distribution
of orbital magnetization corresponding to the quasiparticle energy renormalization.
\end{abstract}


\maketitle

\section{Introduction}\label{sec:intro}
Magnetization is one of the most important and intriguing material properties. An
adequate account of magnetization should not only include the
contribution from the spin polarization of electrons, but also the
contribution from the orbital motion of electrons. In crystals, due
to the reduced spatial symmetry, the orbital contribution to the
magnetization is usually quenched. However, in certain materials
with topologically nontrivial band structures, large contributions
can arise from the effective reciprocal space monopoles near the band anti-crossings. Several
different methods have been employed to study the orbital
magnetization (OM) in crystals~\cite{Hirst,Xiaod,Thonhauser11,Xiao,Yao,Thonhauser,Ceresoli06,
Souza,Shi,Ceresoli07,Ceresoli,Resta,Restab,Gata,Gatb,wanga,wangb}.
One major difficulty in the calculation is posed by the evaluation of the
operator~\cite{Hirst,Xiaod} $\hat{\bm r}\times\hat{\bm j}$, because the
the position $\hat{\bm r}$ is ill-defined in the Bloch
representation. This difficulty can be avoided in a semiclassical picture or be circumvented by a transformation to
the Wannier representation. Xiao {\it et al.}~\cite{Xiao,Yao} presented a general formula for OM for metal and insulator,
derived from a semiclassical formalism with the
Berry phase corrections. Thonhauser {\it et al.}~\cite{Thonhauser,Ceresoli06}
derived an expression of the OM for periodic insulators using the Wannier representation.
From the elementary thermodynamics, Shi {\it et al.}~\cite{Shi} obtained a formula for the OM in a periodic
system using the standard perturbation theory. Their result can in principle take into account the electron-electron interaction effects.
A computation of the OM for periodic systems with density-functional theory was carried out by Ceresoli {\it et al.}~\cite{Ceresoli}.

Previous studies are mainly concerned with clean systems. However, real crystals are never perfect, disorders
such as defects, impurities, phonons etc. constantly break the translational symmetry and lead to scattering events.
The effect of disorder scattering on the OM has not been carefully studied so far.
On one hand, the OM is a thermodynamic quantities,
hence it is expected to be less susceptible to disorder
scattering. On the other hand, the appearance of current operator $\hat{\bm j}$ in the definition
suggests behaviors similar to transport quantities which might be strongly affected by
the disorder scattering. Therefore, it is important and desirable to have a good understanding of
the role played by the disorder scattering in the OM.

In this paper, we present a general formula of the OM in
disordered systems based on the Keldysh Green's function
theory in the gauge-covariant Wigner space~\cite{Onoda06,Onoda,Onoda08,Sinova,Haug}. This approach was developed as a systematic approach to
the nonequilibrium electron dynamics under external fields. Our formula derived from this approach
shares the advantage of being able to capture the disorder effects in a systematic way and
ensure the gauge invariance property from the very beginning. We show that in the clean limit,
our formula reduces to the previous results obtained from other approaches. As an application, we study the
OM in a disordered two-dimensional electron gas with the Rashba spin-orbit coupling.
We find that the OM is robust against short range disorders. The main effect of the
scattering by short range disorders is a rigid shift of the distribution of OM in energy.

The structure of this paper is organized as follows. In
Sec.~\ref{sec:gfinwiger}, we outline the Keldysh Green's
function formalism which is employed for our derivation. Our general formula
of OM is presented in Sec.~\ref{sec:ominwiger}. In
Sec.~\ref{sec:apply}, we apply the formula to study the OM of a
two-dimensional disordered electron gas with the Rashba spin-orbit coupling.
Summary and conclusion are given in
Sec.~\ref{sec:conclusions}.
Some details of the calculation are provided in the appendices.

\section{Orbital magnetization of disordered systems}\label{sec:modmet}

\subsection{Keldysh Green's function formalism} \label{sec:gfinwiger}

We employ the Keldysh Green's function
formalism in the Wigner representation~\cite{Onoda06}, which has recently been used
to study the current response of multi-band systems under an electric field~\cite{Onoda,Onoda08}.
In the Wigner representation, Green's functions and the
self-energies are expressed as functions of the
center-of-mass coordinates ($T$,$\bm{X}$), the
energy $\varepsilon$ and the mehanic momentum $\textbf{p}$. The
energy and the mechanic momentum are the Fourier transforms of the relative time and space coordinates
respectively.

The Dyson equations in the presence of external electromagnetic fields can be written as
\begin{subequations}
  \begin{eqnarray}
    \left[\varepsilon\underline{\hat{I}}-\underline{\hat{H}}_0(\textbf{p})-\underline{\hat{\Sigma}}
    (\varepsilon)\right]\star\underline{\hat{G}}(\varepsilon,\textbf{p})&=&\underline{\hat{I}},
      \label{eq:_Dyson_:1}\\
      \underline{\hat{G}}(\varepsilon,\textbf{p})\star\left[\varepsilon\underline{\hat{I}}-
      \underline{\hat{H}}_0(\textbf{p})-\underline{\hat{\Sigma}}(\varepsilon)\right]&=&\underline{\hat{I}}.
      \label{eq:_Dyson_:2}
  \end{eqnarray}
  \label{eq:Dyson}
\end{subequations}
Each quantity with an underline in the above equations is a matrix in Keldysh space. Specifically, we have
\begin{align}
  \begin{matrix}
   \underline{\hat{G}}&\equiv&\left(\begin{array}{cc}
      \hat{G}^R & 2\hat{G}^<\\
      0         &  \hat{G}^A
    \end{array}\right),
  \end{matrix}
   \quad
  \begin{matrix}
    \underline{\hat{\Sigma}}&\equiv&\left(\begin{array}{cc}
      \hat{\Sigma}^R & 2\hat{\Sigma}^<\\
      0         &  \hat{\Sigma}^A
    \end{array}\right),
  \end{matrix}
\end{align}
\begin{align}
  \begin{matrix}
    \underline{\hat{H}}_0&\equiv&\left(\begin{array}{cc}
      \hat{H}_0 & 0 \\
      0         &  \hat{H}_0
    \end{array}\right),
  \end{matrix}
   \quad
  \begin{matrix}
    \underline{\hat{I}}&=&\left(\begin{array}{cc}
      \hat{\sigma}^0 & 0 \\
      0 & \hat{\sigma}^0
    \end{array}\right),
    \end{matrix}
\end{align}
where $\hat{G}^{(R,A,<)}$ are the (retarded, advanced, lesser) Green functions, and
$\hat{\Sigma}^{(R,A,<)}$ are the corresponding self-energies, $\hat{H}_0$ is the Hamiltonian
in the absence of external electromagnetic fields,
$\hat{\sigma}^0$ is the identity matrix. The $\star$ operator in Eq.(\ref{eq:Dyson}) is defined as
\begin{equation}
  \star\equiv\exp\left[\frac{iq\hbar}{2}F^{\mu\nu}\left(\overleftarrow{\partial}_{p^\mu}\overrightarrow{\partial}_{p^\nu}
  -\overleftarrow{\partial}_{p^\nu}\overrightarrow{\partial}_{p^\mu}\right)\right],
  \label{eq:Moyal}
\end{equation}
with the differential operators $\overleftarrow{\partial}$ and
$\overrightarrow{\partial}$ operating on the left-hand and the
right-hand sides respectively, $\bm{q}=-\left\vert e\right\vert$ is
the electron charge, and
$F^{\mu\nu}=\partial_{X_\mu}A^\nu(X)-\partial_{X_\nu}A^\mu(X)$ is
the electromagnetic field tensor, $\mu$ and $\nu$ label the four dimensional space-time components
and the Einstein summation convention is assumed. It should be noted that the
energy $\varepsilon$ and the mechanic momentum $\textbf{p}$
include the electromagnetic potentials $A^\mu(X)$, both are gauge invariant quantities.
The $\star$ operator in Eq.(\ref{eq:Dyson}) only involves the physical fields, so it is also gauge invariant.
In this formalism the gauge invariance is respected from the very beginning and
easily maintained during the perturbative expansion, which is an important advantage~\cite{Onoda06}.

Here we consider the situation with a uniform weak magnetic field along the z-direction, i.e. $\textbf{B}=(0,0,B)$.
Then the various quantities can be expanded in terms of $B$.
In particular, Green's functions and the self-energies can be expressed as
\begin{eqnarray}
  \hat{G}^\alpha(\varepsilon,\textbf{p}) &=& \hat{G}^\alpha_0(\varepsilon,\textbf{p})
+ e\hbar B \hat{G}^\alpha_{B}(\varepsilon,\textbf{p})+O(B^2),
  \label{eq:g}\\
  \hat{\Sigma}^\alpha(\varepsilon) &=& \hat{\Sigma}^\alpha_0(\varepsilon)
+ e\hbar B \hat{\Sigma}^\alpha_{B}(\varepsilon)+O(B^2),
  \label{eq:Sigma}
\end{eqnarray}
with $\alpha=R,A,<$ for the retarded, advanced and lesser
components respectively.
Here functions with the subscript 0 are of zeroth order in the external
magnetic field strength (note that they include scattering effects). We have
\begin{eqnarray}
  \hat{G}^{R(A)}_0(\varepsilon,\textbf{p})=\left[\varepsilon-\hat{H}_0(\textbf{p})
-\hat{\Sigma}^{R(A)}_0(\varepsilon)\right]^{-1},
  \label{eq:G^R,A:0}\\
   \hat{G}^<_0(\varepsilon,\textbf{p})=\left[\hat{G}^A_0(\varepsilon,\textbf{p})-
  \hat{G}^R_0(\varepsilon,\textbf{p})\right]f(\varepsilon),
  \label{eq:G^<_0}
\end{eqnarray}
where $f(\varepsilon)$ is the Fermi distribution. The functions with subscript $B$
are the linear response coefficient to the external field. They can be solved
from the Dyson equation. It is usually convenient
to decompose the lesser component $\hat{G}^<_{B}$ and $\hat{\Sigma}^<_{B}$
(which are related to particle distribution)
into two parts, with one part from the Fermi surface and the other part from
the Fermi sea~\cite{Streda},
\begin{eqnarray}
&&\hat{G}^<_{B}(\varepsilon,\textbf{p})\!\!=
\hat{G}^<_{B,I}(\varepsilon,\textbf{p})\partial_\varepsilon
f(\varepsilon)\!\!+\!\!\hat{G}^<_{B,II}(\varepsilon,\textbf{p})f(\varepsilon)
\label{eq:G^<:b}\\
  &&\hat{\Sigma}^<_{B}(\varepsilon)=\hat{\Sigma}^<_{B,I}(\varepsilon)
\partial_\varepsilon f(\varepsilon)+\hat{\Sigma}^<_{B,II}(\varepsilon)
f(\varepsilon).
  \label{eq:Sigma^<:b}
\end{eqnarray}
From the Dyson equation (kept to the linear order in $B$), it is straightforward to show that
\begin{equation}
\hat{G}^<_{B,I} =\hat{\Sigma}^<_{B,I}=0, \label{eq:G}\\
\end{equation}
i.e. there is no Fermi surface term in the linear order lesser component, and for the Fermi sea term we have
\begin{eqnarray}
 &&\hat{G}^<_{B,II}(\varepsilon,\textbf{p})=\hat{G}^A_{B}(\varepsilon,\textbf{p})
-\hat{G}^R_{B}(\varepsilon,\textbf{p}),\label{eq:G^<:b,II}\\
  &&\hat{\Sigma}^<_{B,II}(\varepsilon)=\hat{\Sigma}^A_{B}(\varepsilon)
-\hat{\Sigma}^R_{B}(\varepsilon).
  \label{eq:Sigma^<:b,II}
\end{eqnarray}
The retarded and advanced Green's function $\hat{G}^{R(A)}_{B}$ and self-energy
$\hat{\Sigma}^{R(A)}_{B}$ are
determined from the following self-consistent equations
\begin{eqnarray}
  \hat{G}^{R(A)}_{B}&=&\frac{i}{2}\left[\hat{G}^{R(A)}_0\hat{v}_x(\partial_{p_y}\hat{G}^{R(A)}_0)
      -(\partial_{p_y}\hat{G}^{R(A)}_0)\hat{v}_x\hat{G}^{R(A)}_0\right]
      \nonumber\\
      &&+\hat{G}^{R(A)}_0\hat{\Sigma}^{R(A)}_{B}\hat{G}^{R(A)}_0,
  \label{eq:G}
\end{eqnarray}
where the velocity operator is defined as
$\hat{v}_i\equiv\frac{1}{i\hbar}\left[\hat{x}_i,\hat{H}\right]$.

In this approach, the disorder effects are captured by the self-energies
$\hat{\Sigma}^{R(A)}_0$ and $\hat{\Sigma}^{R(A)}_{B}$, which allows a systematic
perturbative treatment. In the weak disorder regime, the
self-consistent $T$-matrix approximation provides a good approximation scheme. In this approximation, we have
\begin{eqnarray}
  \hat{\Sigma}^{R(A)}_0(\varepsilon)&=&n_{\rm{imp}}\hat{T}^{R(A)}_0(\varepsilon),
  \label{eq:Sigma^R,A:0}
\end{eqnarray}
and
\begin{eqnarray}
 \hat{\Sigma}^{R(A)}_{B}(\varepsilon)\!&=&\!n_{\rm imp}
\hat{T}^{R(A)}_0(\varepsilon)\!\!\int\!\!\!\frac{d^2\textbf{p}}{(2\pi\hbar)^2}\!
\hat{G}^{R(A)}_{B}(\varepsilon,\textbf{p})\hat{T}^{R(A)}_0(\varepsilon)\!,
\nonumber\\
  \label{eq:Sigma^R,A:b}
\end{eqnarray}
where $n_{\rm
imp}$ is the impurity concentration and the $T$-matrix is expressed as
\begin{eqnarray}
  \hat{T}^{R(A)}_0(\varepsilon)&=&\hat{V}_{\text{imp}}\left(1-\int
\frac{d^2\textbf{p}}{(2\pi\hbar)^2}\hat{G}^{R(A)}_0(\varepsilon,\textbf{p})\hat{V}_{\text{imp}}\right)^{-1},
\nonumber\\
  \label{eq:g^R,A:0}
\end{eqnarray}
with $\hat{V}_{\text{imp}}$ being the impurity potential.

The equilibrium Green's functions $\hat{G}^{R(A)}_0$ and
the self energies $\hat{\Sigma}^{R(A)}_0$ can be obtained by solving
Eqs.~(\ref{eq:G^R,A:0}), ~(\ref{eq:Sigma^R,A:0}) and~
(\ref{eq:g^R,A:0}) self-consistently. Then the linear order coefficients $\hat{G}^{R(A)}_{B}$ and
$\hat{\Sigma}^{R(A)}_{B}$ can be solved from Eqs.~(\ref{eq:G}) and~
(\ref{eq:Sigma^R,A:b}). Finally, we can obtain $\hat{G}^<_{B,II}$ through
Eq.~(\ref{eq:G^<:b,II}) and the linear response of the system in the external magnetic field can be
completely determined.

The lesser Green's function contains the information of particle distribution. In our case, both the external magnetic field and the disorder scattering affect the quasiparticle distribution.
Before we proceed, it is interesting to observe how the non-trivial band geometry (described by the Berry curvature)
can be captured by the present Wigner space Green's function formalism. For a homogeneous system, the electron density can be
written as
\begin{equation}
\bm n_e = \frac{1}{i}\int\frac{d\varepsilon}{2\pi}
\int\frac{d^{2}\textbf{p}}{(2\pi\hbar)^{2}} \mathrm{tr}\left[
\hat{G}^{<}(\varepsilon,\textbf{p})\right].
\label{eq:number}
\end{equation}
In the absence of the disorder scattering, the eigenstates are well-defined Bloch states grouped into energy bands.
Using the theorem of residues, we can express the ground state electron density
in the presence of a constant magnetic field as
(see Appendix \ref{edensity})
\begin{eqnarray}\label{density}
\bm n_e &=& \sum_{n,occ}\int\!\frac{d^2\textbf{p}}{(2\pi\hbar)^2}\left[ 1 + \frac{e}{\hbar}\textbf{B}\cdot\bm\Omega_{n}(\textbf{p})\right]. \label{eq:ed}
\end{eqnarray}
The summation is over all the occupied states, and $\bm\Omega_{n}(\textbf{p}) \!=\!
i\langle{\bm\nabla_{\textbf{p}}u_{n\textbf{p}}|\times|\bm\nabla_{\textbf{p}}u_{n\textbf{p}}}\rangle$ is the Berry curvature
of the Bloch state $|n,\textbf{p}\rangle=e^{i\textbf{p}\cdot\textbf{x}/\hbar}|u_{n\textbf{p}}\rangle$.
It can be seen that the Fermi-sea volume is
changed linearly by a magnetic field when the Berry curvature is nonzero. This effect was previous interpreted
as the modification of phase space density of states~\cite{Xiao}.

\subsection{Formula of orbital magnetization} \label{sec:ominwiger}

We start from the standard thermodynamic definition of the OM density at zero temperature~\cite{Shi}:
\begin{equation}
\bm M =-\left(\frac{\partial K}{\partial\bm B}\right)_\mu,
  \label{eq:om}
\end{equation}
where $K=E-\mu N$ is the grand thermodynamic potential, $B$ is a weak magnetic field.
Since we are concerned with the orbital contribution, the small Zeeman coupling between the electron spin
and external field will be dropped.
The potential $\bm K$ can be expressed through the lesser Green's function,
\begin{equation}
\bm K = \frac{1}{i}\int\frac{d\varepsilon}{2\pi}
\int\frac{d^{2}\textbf{p}}{(2\pi\hbar)^{2}} \mathrm{tr}\left[
\left(\hat{H}-\mu \right) \hat{G}^{<}(\varepsilon,\textbf{p})\right].
\label{eq:Me}
\end{equation}
Using Eqs.~(\ref{eq:om}),
~(\ref{eq:Me}) and ~(\ref{eq:G^<:b}), we find that the OM can be written as
\begin{eqnarray}
  \bm M&=&-ie\hbar\int\frac{d\varepsilon}{2\pi}
  \int\frac{d^2\textbf{p}}{(2\pi\hbar)^2}\mathrm{tr}\left[
  \left(\hat{H}-\mu\right)
  \right. \nonumber\\
  &&\times\left.\left(\hat{G}^A_B(\varepsilon,\textbf{p})-
  \hat{G}^R_B(\varepsilon,\textbf{p})\right)\right]f(\varepsilon).
  \label{eq:om_II}
\end{eqnarray}
From this expression, we can see that the OM has contributions from the
whole Fermi sea, with no separate Fermi surface contribution such as that for the transport quantities.

In this formula, the impurity scattering effect comes in through two terms:
the self-energy $\hat{\Sigma}^{R,A}_0$ which modifies the ground state electronic structure and the vertex
corrections associated with $\hat{\Sigma}^{R,A,<}_B$ which represent
an interplay between the magnetic field and the impurity scattering. We may separate out the terms
containing $\hat{\Sigma}^{R,A,<}_B$ and write the OM explicitly as
\begin{equation}
  \bm M=\bm M^{I}+\bm M^{II},
  \label{eq:sigma_xy^Int}
\end{equation}
where
\begin{eqnarray}
  \bm M^{I}
  =\frac{e\hbar}{2}&&\int\!\frac{d\varepsilon}{2\pi}f(\varepsilon)\int\!\frac{d^2\textbf{p}}{(2\pi\hbar)^2}
  \nonumber\\
  \times
  \sum_{ij}\text{Tr}&&\left[\epsilon_{ij}(\hat{H}-\mu)\hat{G}^A_0(\varepsilon,\textbf{p})\hat{v}_i\hat{G}^A_0(\varepsilon,\textbf{p})\hat{v}_j
  \hat{G}^A_0(\varepsilon,\textbf{p})
    \right.\nonumber\\
    &&\left.-(\hat{G}^A_0\rightarrow \hat{G}^R_0)\right],
  \label{eq:IOM:int}
\end{eqnarray}
and
\begin{eqnarray}
  &&\bm M^{II}
  =-ie\hbar\int\!\frac{d\varepsilon}{2\pi}f(\varepsilon)\int\!\frac{d^2\textbf{p}}{(2\pi\hbar)^2}
  \nonumber\\
  &&\times\text{Tr}\left[(\hat{H}-\mu)\hat{G}^{A}_0\hat{\Sigma}^{A}_B\hat{G}^{A}_0-(\hat{H}-\mu)
  \hat{G}^{R}_0\hat{\Sigma}^{R}_B\hat{G}^{R}_0\right],
  \label{eq:IOM:ext}
\end{eqnarray}
where $\epsilon_{ij}$ with $i,j\in\{x,y\}$ is the 2D antisymmetric tensor, and
the second term in the bracket in Eq.(\ref{eq:IOM:int}) means that the second term is the same as the first term except
that all the $\hat{G}^A_0$ are replaced by $\hat{G}^R_0$.
Such a decomposition scheme was also adopted in the study of anomalous Hall conductivity~\cite{Onoda08},
and in that context, the two parts are referred to as the intrinsic part
and extrinsic part respectively. It should be noted that the intrinsic part $\bm M^{I}$
also has impurity scattering effects in it (see Eq.~(\ref{eq:G^R,A:0}) and Eq.~(\ref{eq:Sigma^R,A:0})), it is
intrinsic in the sense that it only contains quantities that are of zeroth order in the external field.
As for the extrinsic part $\bm M^{II}$, it is easy to see that it is already linear order
in $n_{\rm imp}$ (see Eq.~(\ref{eq:Sigma^R,A:b})). Therefore in the weak scattering regime, the extrinsic part is
expected to be much smaller than the intrinsic part.

The above formula is our main result. From this formula, we see that there is no separate
Fermi surface contributions like those in the transport quantities, which is consistent with
OM being a thermodynamic equilibrium property.
This formula applies for both insulators and metals. The quantities in this
formula can be calculated from the Dyson equation according to our prescription described in the previous section.
It can also be straightforwardly implemented in the numerical calculation, either from effective models or from first principles.

In the clean limit, we only have the intrinsic part. The general result reduces to (see Appendix \ref{integralom}
for the derivation)
\begin{eqnarray}\label{cleanM}
\bm M &=& \sum_{n\textbf{p}}f_{n\textbf{p}}\left[ \bm m_{n}(\textbf{p})
 - \frac{e}{\hbar}(\epsilon_{n\textbf{p}}-\mu)\bm\Omega_{n}(\textbf{p})\right],
 \label{eq:MT-final2}
\end{eqnarray}
where
$\bm m_{n}(\textbf{p})\!=\!(e/2\hbar)i\langle\bm\nabla_{\textbf{p}}u_{n\textbf{p}}|
[\epsilon_{n}(\textbf{p})-\hat{H}_{0}(\textbf{p})]\times|\bm\nabla_{\textbf{p}}u_{n\textbf{p}}\rangle$ is the orbital moment
of the Bloch state $|n, \textbf{p}\rangle$ and $\bm\Omega_{n}(\textbf{p}) \!=\!
i\langle{\bm\nabla_{\textbf{p}}u_{n\textbf{p}}|\times|\bm\nabla_{\textbf{p}}u_{n\textbf{p}}}\rangle$ is the Berry curvature.
The first term in Eq.~(\ref{eq:MT-final2}) is a sum of the orbital magnetic moments associated with each Bloch state~\cite{Chang,Sundaram},
and the second term is a Berry-phase correction to the OM. Therefore, the OM can be written as
\begin{eqnarray}\label{cleanM}
\bm M &=& \mathcal{M}_{\text{m}}+\mathcal{M}_{\Omega}.
\label{eq:MT-final2b}
\end{eqnarray}
This clean limit result was previously derived
from the standard perturbation theory of quantum mechanics
by Shi {\it et al.} ~\cite{Shi} and also from the semiclassical theory by Xiao
{\it et al.} ~\cite{Xiao}. Now it is also reproduced as a special limiting case of our general formula.

\section{Application to a two-dimensional electron gas with
Rashba spin-orbit coupling}\label{sec:apply}

\subsection{Model}\label{sec:model}

\begin{figure}

 \includegraphics[width=8.0cm]{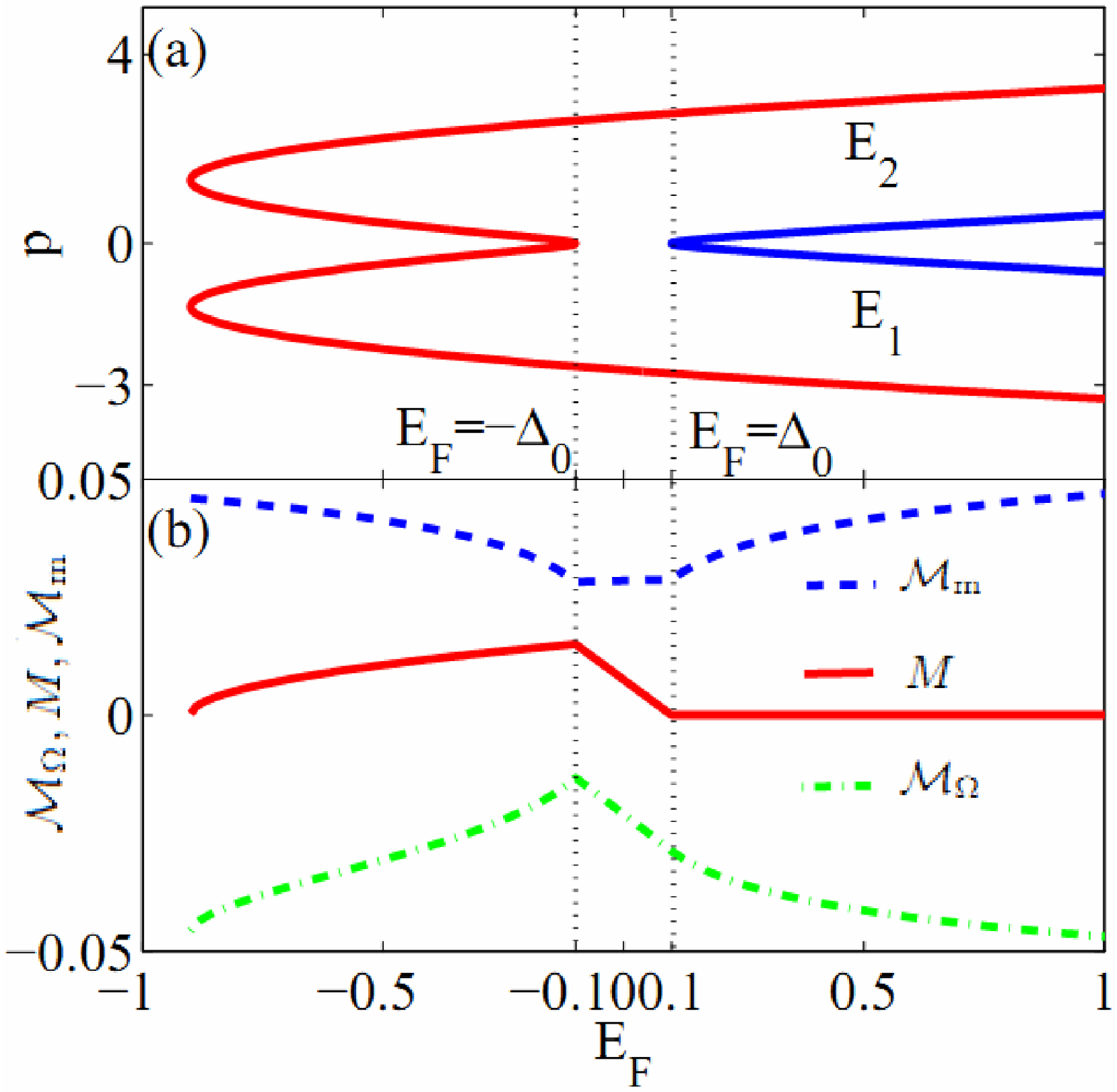}
  \caption{(Color online) (a) Electronic band dispersions of our model given by Eq.~(\ref{eq:energy}). (b) Orbital magnetization ($\bm M$, solid red curve) of disordered free system and its two components $\mathcal{M}_\text{m}$ (dashed blue curve) and $\mathcal{M}_\Omega$ (dash-dotted green curve) as functions of Fermi energy $E_F$. They are plotted in units of $e/{\hbar}$. The parameters are chosen as $2m\alpha^2=3.59$ and $\Delta_0=0.1$.}
    \label{fig:OMth1}
\end{figure}

\begin{figure}

 \includegraphics[width=8.0cm]{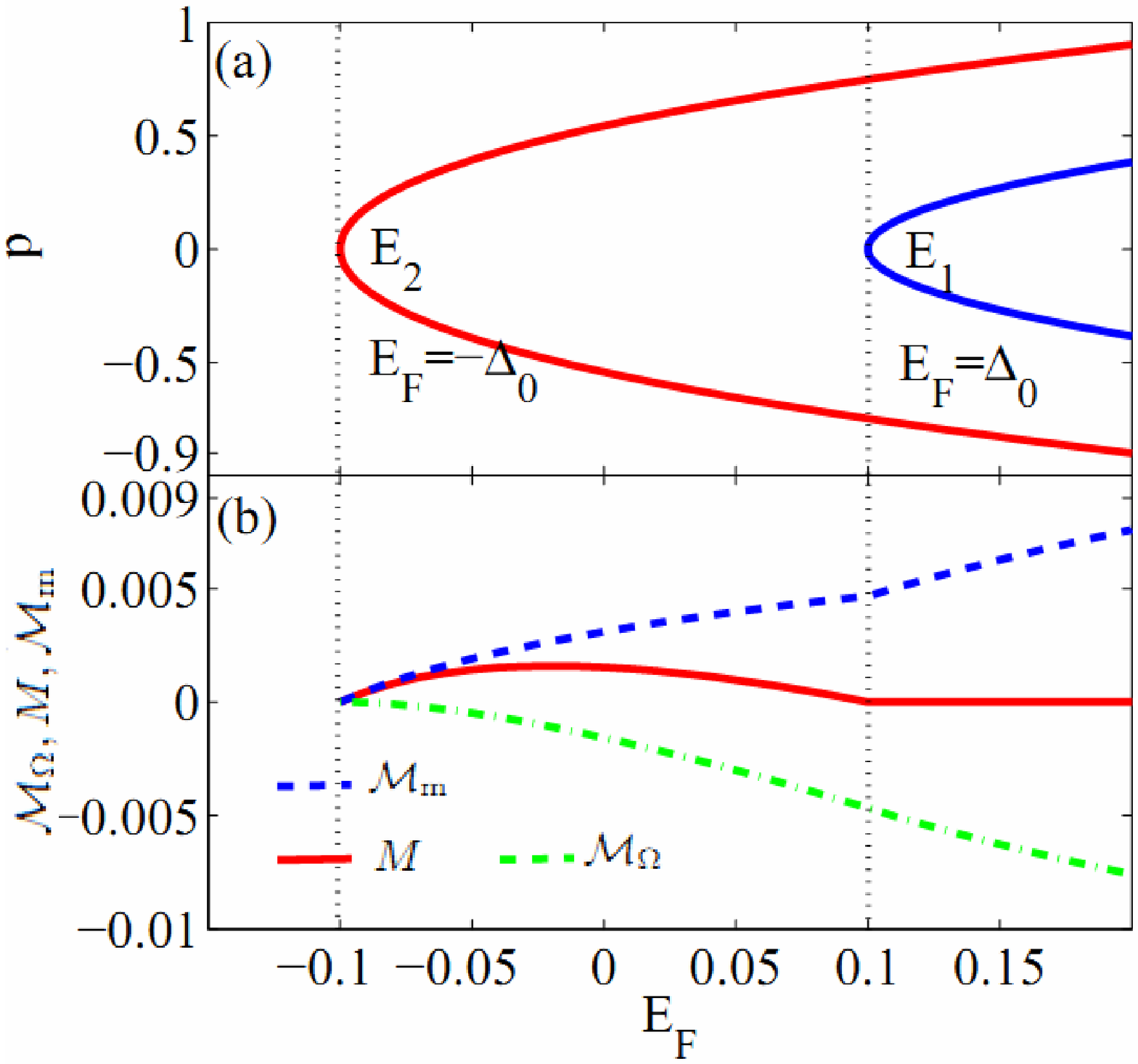}
  \caption{(Color online) (a) Electronic band dispersions of our model given by Eq.~(\ref{eq:energy}). (b) Orbital magnetization ($\mathcal{M}$, solid red curve) of disorder free system and its two components $\mathcal{M}_\text{m}$ (dashed blue curve) and $\mathcal{M}_\Omega$ (dash-dotted green curve) as functions of Fermi energy $E_F$. They are plotted in units of $e/{\hbar}$. The parameters are chosen as $2m\alpha^2=0.08$ and $\Delta_0=0.1$.}
    \label{fig:OMth2}
\end{figure}

We apply our theory to study the model of a two-dimensional disordered electron gas with
Rashba spin-orbit coupling. The Hamiltonian for the system reads
\begin{subequations}
  \begin{eqnarray}
    \hat{H} &=& \hat{H}_0+\hat{H}_\text{imp},
    \label{eq:H}\\
    \hat{H}_0
    &=&\frac{p^2}{2m}\hat{\sigma}^0+\alpha\left(p_x\hat{\sigma}^y-p_y\hat{\sigma}^x  \right)-\Delta_0\hat{\sigma}^z,
    \label{eq:H_0}\\
    \hat{H}_\text{imp} &=& u_{\text{imp}}\hat{\sigma}^0\sum_{\vec{r}_{\rm imp}}\delta(\vec{r}-\vec{r}_{\rm imp}),
    \label{eq:H_imp}
  \end{eqnarray}
  \label{eq:Hs}
\end{subequations}
where $(\hat{\sigma}^x,\hat{\sigma}^y,\hat{\sigma}^z)$
are the three Pauli matrices and $\hat{\sigma}^0$ is the
identity matrix, $\alpha$ is the strength of the
spin-orbit coupling, and term $-\Delta_0\hat{\sigma}^z$ is the spin splitting
which can be introduced by the exchange coupling with a nearby ferromagnet
or magnetic dopants.
$\hat{H}_\text{imp}$ is the disorder potential from the
randomly distributed short range impurities with strength
$u_{\text{imp}}$. The energy
dispersion of the Hamiltonian $\hat{H}_0$ is given by
\begin{eqnarray}
\bm E_\lambda(p)=\frac{p^2}{2m}-(-1)^\lambda\sqrt{\Delta^2_0+\alpha^2p^2},
\label{eq:energy}
\end{eqnarray}
where $\lambda=1,2$ labels the upper and lower band respectively. When the Rashba coupling energy scale~\cite{Garelli} $m\alpha^2$ is larger
than the Zeeman coupling strength the minima of the lower band occur at a finite wave vector
and the dispersion assumes a Mexican hat shape (see Fig.~\ref{fig:OMth1} (a)). When the Zeeman coupling dominates over
the Rashba energy, the minimum of the lower band $\bm E_2$ occurs at the origin (see Fig.~\ref{fig:OMth2} (a)).


Let's first consider the clean limit,
in which case the orbital magnetic moment and the Berry curvature of each Bloch state
can be calculated straightforwardly:
\begin{eqnarray}
\bm m_{1}(\bm p)=\bm m_{2}(\bm
p)=\frac{e}{2\hbar}\frac{\Delta_0\alpha^2}{\Delta^2_0+\alpha^2p^2},
\label{eq:omclean}\\
\bm\Omega_{1}(\bm p)=-\bm\Omega_{2}(\bm
p)=-\frac{1}{2}\frac{\Delta_0\alpha^2}{(\Delta^2_0+\alpha^2p^2)^\frac{3}{2}}.
\label{eq:bcclean}
\end{eqnarray}
It is interesting to observe that for the same wavevector the orbital moments
of the two bands have the same magnitude and the same sign,
while the Berry curvatures
have the same magnitude but opposite signs. It should also be noted that both the orbital
moment and the Berry curvature would vanish if either $\alpha$ or
$\Delta_0$ vanishes. From Eq.(\ref{eq:MT-final2}), we further see that the OM is
\emph{nonzero} only when both the spin-orbit coupling and the exchange coupling are present.

Analytical expressions of the OM can be easily
obtained for the clean limit using Eq.(\ref{eq:MT-final2}). For example, for the case with $E_F>\Delta_0$,
we have

\begin{eqnarray}
\bm M&=&\frac{e\Delta_0}{4\pi\hbar}(E_F\!+\!\frac{\Delta^2_0}{2m\alpha^2})\left[\frac{1}{(\Delta^2_0
+\alpha^2p_{F_1}^2)^\frac{1}{2}}\!-\!\frac{1}{(\Delta^2_0+\alpha^2p_{F_2}^2)^\frac{1}{2}}\right]
\nonumber\\
&&+\frac{e\Delta_0}{8\pi m\hbar \alpha^2}[(\Delta^2_0
+\alpha^2p_{F_1}^2)^\frac{1}{2}-(\Delta^2_0
+\alpha^2p_{F_2}^2)^\frac{1}{2}],
   \label{eq:analytical}
\end{eqnarray}
where $p_{F_{1,2}}$ is the Fermi momenta of the two bands.

\subsection{Results}\label{sec:result}

\begin{figure}
 \includegraphics[width=8.0cm]{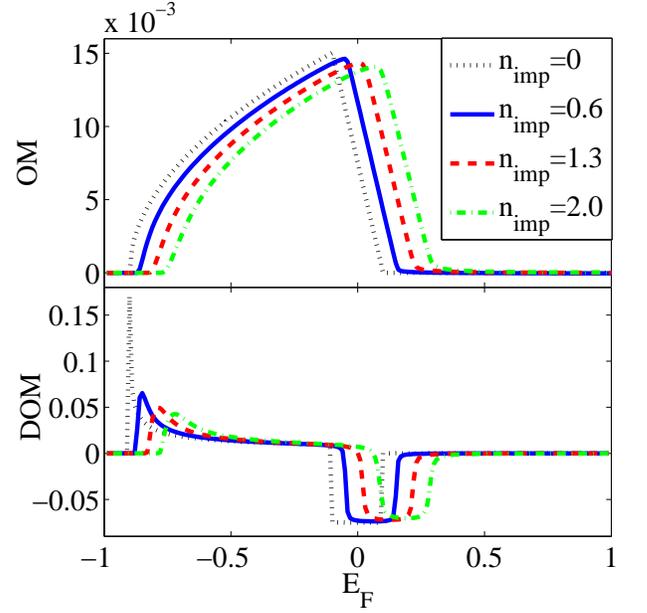}
  \caption{(Color online) (a) Orbital magnetization (OM) as a function of Fermi energy $E_F$ with different impurity concentration $n_{\text{imp}}$.  (b) Density of OM with different impurity concentration $n_{\text{imp}}$.  These quantities are plotted in units of $e/{\hbar}$. The parameters are chosen as $2m\alpha^2=3.59$, $\Delta_0=0.1$, and $u_{\text{imp}}=0.1$.}
    \label{fig:IOM1}
\end{figure}

Now we analyze the OM of the disordered 2D Rashba model in detail.
The calculation procedure follows our discussion in Sections ~\ref{sec:gfinwiger} and ~\ref{sec:ominwiger}.
Since we have seen that both the spin-orbit coupling and the exchange coupling are essential ingredient for
the OM, in the following we shall consider two different regimes of the model determined
by the competition between the
Rashba spin-orbit coupling and the exchange coupling. For each regime, we first analyze the clean limit
where the physical picture is more transparent, and then study the influence of disorder scattering which is the focus
in this paper.

\begin{figure}
 \includegraphics[width=8.0cm]{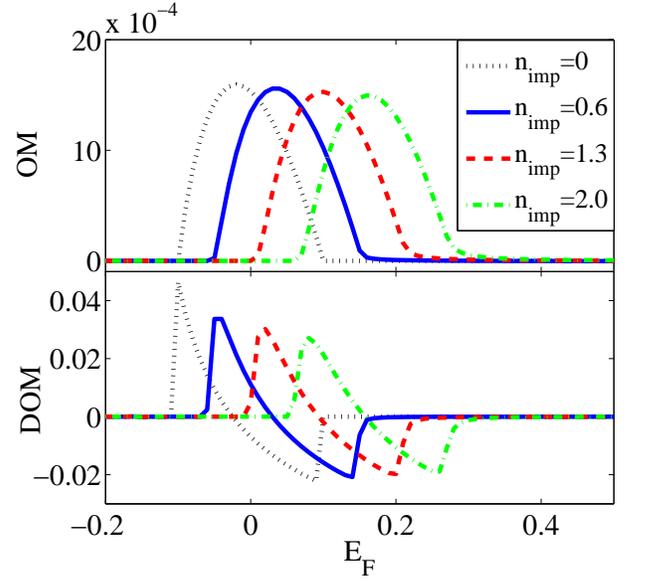}
  \caption{(Color online) (a) Orbital magnetization (OM) as a function of Fermi energy $E_F$ with different impurity concentration $n_{\text{imp}}$. (b) Density of OM with different impurity concentration $n_{\text{imp}}$. These quantities are plotted in units of $e/{\hbar}$. The parameters are chosen as $2m\alpha^2=0.08$, $\Delta_0=0.1$, and $u_{\text{imp}}=0.1$.}
    \label{fig:IOMso1}
\end{figure}

We first consider the regime where the Rashba coupling dominates over the exchange coupling, i.e. $2m\alpha^2\gg \Delta_0$.
The typical band dispersion in this regime is shown in Fig.~\ref{fig:OMth1} (a) (with $2m\alpha^2=3.59$ and $\Delta_0=0.1$).
In this regime, the bottom of the lower band occurs at a finite wavevector.
The energy spectrum around the origin has an
effective Dirac cone
structure with a local gap $2\Delta_0$ at $p=0$. Both the orbital moment and the Berry curvature are
concentrated near this band anticrossing point, as is evident from Eqs.(\ref{eq:omclean}) and (\ref{eq:bcclean}).
Fig.~\ref{fig:OMth1} (b) shows the OM
for the clean limit. The orbital moment contribution
$\mathcal{M}_\text{m}$ and the Berry curvature contribution $\mathcal{M}_\Omega$ are also plotted in Fig.~\ref{fig:OMth1} (b).
We can see that as the Fermi energy $E_F$ increases from the lower band bottom, $\mathcal{M}_\Omega$ increases while $\mathcal{M}_\text{m}$
decreases. The increasing rate of $\mathcal{M}_\Omega$ is higher than the decreasing rate of
$\mathcal{M}_\text{m}$, so the overall OM is increasing.
The OM reaches its maximum when $E_F=-\Delta_0$, which corresponds to the
local band top around the origin in momentum space. As the Fermi energy sweeps across the local energy gap between $-\Delta_0$ and $+\Delta_0$, the OM decreases approximately linearly with $E_F$. The linearity can be understood by noticing that
from Eq.(\ref{eq:MT-final2})) the derivative of the OM with respect to $E_F$ is
just the momentum space integral of the Berry curvature. The Berry curvature distribution is concentrated near the
band anticrossing point, corresponding to the small region around the origin in the present model. When the Fermi energy
is within the gap, the Berry curvature integral only has contribution from the lower band and is almost constant, therefore
leading to the linear energy dependence of OM.
This linear decrease of OM stops when the Fermi energy touches the bottom of the upper band at $+\Delta_0$.
Above the upper band bottom, $\mathcal{M}_\Omega$ and $\mathcal{M}_\text{m}$ almost cancel each other and the OM is vanishingly small.
Throughout the spectrum, $\mathcal{M}_\text{m}$ is positive while $\mathcal{M}_\Omega$ is negative, corresponding to the paramagnetic
and diamagnetic responses respectively. This has a clear explanation in the semiclassical picture: $\mathcal{M}_\text{m}$ is due to the self-rotation of the wavepacket which is paramagnetic, while $\mathcal{M}_\Omega$ is from the center-of-mass motion
of the wavepacket hence is diamagnetic~\cite{Xiaod}.

When the exchange coupling dominates over the Rashba energy, The minimum of the lower band occurs at
the origin. Compared with the previous case, there is no local gap
at $p=0$. The typical band dispersion is shown in Fig.~\ref{fig:OMth2} (a) (with $\Delta_0=0.1$ and take $2m\alpha^2=0.08$).
The overall shape of the OM is similar to that for the first case. Its distribution over spectrum is mainly below the upper band bottom.
However, due to the absence of the local gap, the kink point at $-\Delta_0$ in Fig.~\ref{fig:OMth1} (a) merges with the
lower band bottom. Moreover, the two contributions $\mathcal{M}_\Omega$ and $\mathcal{M}_\text{m}$
strongly cancel each other and the resulting OM is much smaller.

\begin{figure}
  \begin{center}
    \includegraphics[width=8.0cm]{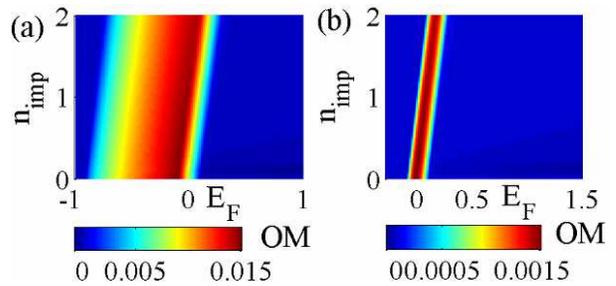}
  \end{center}

 \caption{(Color online) Orbital magnetization (OM) as functions of Fermi energy
    $E_F$ and the impurity concentration $n_{\text{imp}}$ in units of $e/{\hbar}$. The parameters are chosen as $\Delta_0=0.1$, and $u_{\text{imp}}=0.1$, except the Rashba energy: (a) $2m\alpha^2=3.59$, (b) $2m\alpha^2=0.08$.}
  \label{fig:IOM2}
\end{figure}

Now let's consider the effects of disorder scattering on the OM in our model. When the disorder scattering is turned on,
the translational invariance is broken. We can no longer define quantities such as $\mathcal{M}_\Omega$ and $\mathcal{M}_\text{m}$.
Their effects are merged into the sophisticated expression in Eq.(\ref{eq:sigma_xy^Int}). Fig.~\ref{fig:IOM1} (a) and Fig.~\ref{fig:IOMso1} (a) show the OM versus $E_F$ for the two regimes we discussed above.
The different curves in each figure correspond to different impurity concentrations $n_{\text{imp}}$.
Compared with the clean limit where $n_{\text{imp}}=0$, we see that the shape of the OM curve is almost unchanged but mainly
its position is shifted by the scattering.
This behavior is more obvious when we look at the density of OM shown in Fig.~\ref{fig:IOM1} (b) and Fig.~\ref{fig:IOMso1} (b).
For the clean limit, we see that the major contribution to the OM is from the states at the band bottom and
at the local band edge.
The effect of disorder scattering here is to shift the the density of OM distribution in energy.
Such a shift can be understood by noticing that the OM only has the Fermi sea contribution. The main effect
of scattering in Eq.(\ref{eq:sigma_xy^Int}) is the shift of energy arising from the real part of the self-energy correction~\cite{Groth}.
For the short range disorder model, the disorder potential is a constant in momentum space, hence
the self-energy is independent of the state, which results in a rigid energy shift for all the states.
For a general disorder potential, the energy shift would be generally different for different states therefore
the distribution of OM would be distorted. The effects of finite range disorders are currently under investigation.

To leading order, the shift should be linear in the disorder density $n_\text{imp}$.
In Fig.~\ref{fig:IOM2} we plot the OM as a function of $E_F$ and $n_\text{imp}$. The linear dependence
of the energy shift in $n_\text{imp}$ is clearly observed.
Apart from the energy shift, the scattering induced state broadening is manifested as the smoothing of the
peaks of the density of OM, which can be clearly observed in Fig.~\ref{fig:IOM1} (b) and Fig.~\ref{fig:IOMso1} (b).
The peaks of OM are only slightly decreased by the scattering.
This means that the OM carried by the electronic states are quite robust against scattering.

\section{Conclusions}\label{sec:conclusions}

In summary, we have derived a formula of the the OM of disordered electron systems based on the
Keldysh Green's function theory. This approach was developed as a systematic approach to
the nonequilibrium electron dynamics under external fields. In the formula, OM is expressed in terms of the Green's functions
and self-energies, which can be solved from the Dyson equations, and systematic approximation schemes to the disorder effects
can be employed. We find that there is no Fermi surface contribution like in the case of the current response. Our formula applies not only for insulators but also for metallic systems, where the quasiparicle behavior is
usually strongly modified by the disorder scattering. It can also be straightforwardly implemented in the numerical calculation. In the clean
limit, our formula reduces to the previous result obtained from other approaches.
As an application, we calculate the OM of a weakly disordered two dimensional electron gas with Rashba spin-orbit coupling.
The result shows that in the simplest white noise short range disorder model,
the OM is robust against weak scattering and the main effect of scattering is a
rigid shift of the distribution of OM in energy,
which can be attributed to the real part of the self energy.

\begin{acknowledgments}
The authors gratefully thank Junren Shi for useful discussions. Y. Y. was supported by the MOST Project of China (Grants No. 2011CBA00100)
and NSF of China (Grants No. 10974231 and 11174337).
W. M. Liu was supported by the NKBRS of China (Grants No. 2011CB921502 and 2012CB821305) and NSF of China (Grants No. 10934010).
\end{acknowledgments}

\appendix

\section{Self-consistent equation for $\hat{\Sigma}^R_0$ and explicit forms of $\hat{G}^R_0$}
\label{app:sigma^R-0}

The Green functions and self-energies in the absence
of the external fields are obtained from the coupled self-consistent
equations~(\ref{eq:G^R,A:0}), ~(\ref{eq:Sigma^R,A:0}) and~
(\ref{eq:g^R,A:0}). In our model, a direct analytical integration in $\varepsilon$ shows that

\begin{eqnarray}
  \Sigma^{R0}_0(\varepsilon)&=&\frac{n_{\rm imp}u_{\text{imp}}\left(1-u_{\text{imp}}g^{R0}_0(\varepsilon)\right)}
  {(1-u_{\text{imp}}g^{R0}_0(\varepsilon))^2-u_{\text{imp}}^2g^{Rz}_0(\varepsilon)^2},
  \label{eq:Rashba:Sigma^R,A0_0}\\
  \Sigma^{Rz}_0(\varepsilon)&=&\frac{n_{\rm imp}u_{\text{imp}}^2g^{Rz}_0(\varepsilon)}{(1-u_{\text{imp}}g^{R0}_0(\varepsilon))^2-u_{\text{imp}}^2g^{Rz}_0(\varepsilon)^2},
  \label{eq:Rashba:Sigma^R,Az_0}\\
   \Sigma^{Rx}_0(\varepsilon)&=&\Sigma^{Ry}_0(\varepsilon)=0,
   \label{eq:Rashba:Sigma^R,Axy_0}
     \end{eqnarray}
  \begin{eqnarray}
  g^{R0}_0(\varepsilon)&=&\frac{m}{4\pi\hbar^2}
  \sum_\sigma\ln\frac{G^{R}_0(\varepsilon,\Lambda,\sigma)}{G^R_0(\varepsilon,0,\sigma)}
  -m\alpha^2 \tilde{g}^{R}_0(\varepsilon),\ \ \ \ \ \ \ \
  \label{eq:Rashba:g^R,A0_0}\\
  g^{Rz}_0(\varepsilon)&=&(-\Delta_0+\Sigma^{Rz}_0(\varepsilon))\tilde{g}^{R}_0(\varepsilon),
  \label{eq:Rashba:g^R,Az_0}
\end{eqnarray}

and
\begin{eqnarray}
  \tilde{g}^{R}_0(\varepsilon)&=&\frac{m}{4\pi\hbar^2 R^{R}(\varepsilon)}\big[
    \sum_\sigma\sigma\ln\left(\varepsilon-p^2/2m
\right.\big.\nonumber\\
&&\left.\big.+
    \mu-\Sigma^{R0}_0(\varepsilon)+m\alpha^2+\sigma
R^{R}(\varepsilon)\right)\big]_{p=0}^{p=\Lambda},
  \label{eq:Rashba:tg^R,A_0}
\end{eqnarray}
\begin{eqnarray}
  G^{R}_0(\varepsilon,p,\pm)&=&\big(\varepsilon-p^2/2m\!+\!\mu\!-\!\Sigma^{R0}_0(\varepsilon)
  \big.\nonumber\\
&&\big.
\mp\sqrt{\alpha^2p^2+(\!-\!\Delta_0\!+\!\Sigma^{Rz}_0(\varepsilon))^2}\big
)^{-1},
  \label{eq:Rashba:G^R,A_0sigma}\end{eqnarray}
\begin{eqnarray}
  R^{R}(\varepsilon)&=&\left((m\alpha^2)^2+2m\alpha^2(\varepsilon+\mu-\Sigma^{R0}_0(\varepsilon))
   \right.\nonumber\\
&&\left.
 +(-\Delta+\Sigma^{Rz}_0(\varepsilon))^{2}\right)^{\frac{1}{2}},
\nonumber\\
  \label{eq:Rashba:F}
\end{eqnarray}
where $\Lambda$ is the cut-off in momentum
integration, and
\begin{eqnarray}
 \hat{G}^R_0(\varepsilon) &=&  G^{R0}_0(\varepsilon) \hat{\sigma}^{0}+ \sum_{\l=x,y,z} G^{Rl}_0(\varepsilon)\hat{\sigma}^l ,
  \label{eq:ffSigma:4}
\end{eqnarray}
with
\begin{eqnarray}
  G^{R0}_0(\varepsilon,\textbf{p})&=&\left(\varepsilon-p^2/2m+\mu-\Sigma^{R0}_0(\varepsilon)\right)\tilde{G}^R_0(\varepsilon,p),
  \nonumber\\
\label{eq:tG^R_0}\\
 G^{Ri}_0(\varepsilon,\textbf{p})&=&\left(-\alpha\epsilon_{ijz}p_j+\delta_{iz}(-\Delta_0+\Sigma^{Rz}_0(\varepsilon))\right)\tilde{G}^R_0(\varepsilon,p),
  \nonumber\\
  \label{eq:tG^R_0}\\
 \tilde{G}^R_0(\varepsilon,p)&=&(\varepsilon-p^2/2m+\mu-\Sigma^{R0}_0(\varepsilon))^2
\nonumber\\
&&+\alpha^2p^2+(-\Delta_0+\Sigma^{Rz}_0(\varepsilon))^2,
 \label{eq:tbarG^R_0}
\end{eqnarray}
and $\epsilon_{ijl}$ is the anti-symmetric tensor, ($i$, $j$, $l$, $\cdots$)
label the Cartesian components. The same results have been obtained in Ref.~\onlinecite{Onoda08}.

For each $\varepsilon$, the self-energy can be calculated by
iterations which can be performed until the the prescribed accuracy
is reached.

\section{Self-consistent equation for $\hat{G}^R_{B}$ and $\hat{\Sigma}^R_{B}$ and their explicit forms}
\label{app:sigma^R_{B_z}}

The equations for solving the first order corrections $\hat{G}^R_{B}$ and $\hat{\Sigma}^R_{B}$ are
presented here.
Using Eqs.~(\ref{eq:G}) and~(\ref{eq:Sigma^R,A:b}), the retarded
Green's function $\hat{G}^R_{B}$ can be rewritten as
\begin{eqnarray}
 \hat{G}^R_{B}(\varepsilon) &=& G^{R0}_{B}(\varepsilon )\hat{\sigma}^{0}+
 \vec{G}^{R}_{B}(\varepsilon)\cdot\hat{\sigma},
 \label{eq:Rashba:Born_all:Sigma^R,A:E1}
\end{eqnarray}
with
\begin{widetext}
\begin{subequations}
  \begin{eqnarray}
    G^{R0}_{B}(\varepsilon,\textbf{p})
    &=&(G^{R0}_0(\varepsilon,\textbf{p})^2+\vec{G}^{R}_0(\varepsilon,\textbf{p})^2)\Sigma^{R0}_{B}(\varepsilon)+
    2G^{R0}_0(\varepsilon) \vec{G}^{R}_0(\varepsilon,\textbf{p})\cdot\vec{\Sigma}^{R}_{B}(\varepsilon)
    \nonumber\\
    &&{}+\tilde{G}^{R}_0(\varepsilon,\textbf{p})(\partial_{p_x}\hat{H}_0(\textbf{p})
    \times
    \vec{G}^{R}_0(\varepsilon,\textbf{p}))\cdot(\partial_{p_y}\hat{H}_0(\textbf{p})),
  \label{eq:two-band:G^<R,A0:E}\\
  \vec{G}^{R}_{B}(\varepsilon,\textbf{p})
  &=&\tilde{G}^{R}_0(\varepsilon,\textbf{p})\vec{\Sigma}^{R}_{B}(\varepsilon)-\tilde{G}^{R}_0(\varepsilon,\textbf{p})G^{R0}_0(\varepsilon,\textbf{p})
  (\partial_{p_x}\hat{H}_0(\textbf{p}))\times(\partial_{p_y}\hat{H}_0(\textbf{p}))
  \nonumber\\
  &&{}-\tilde{G}^{R}_0(\partial_{p_x}H^0_0(\textbf{p})) \vec{G}^{R}_0(\varepsilon,\textbf{p})\times(\partial_{p_y}\hat{H}_0(\textbf{p}))
  +\tilde{G}^{R}_0(\partial_{p_y}H^0_0(\textbf{p})) \vec{G}^{R}_0(\varepsilon,\textbf{p})\times(\partial_{p_x}\hat{H}_0(\textbf{p}))
  \nonumber\\
  &&{}+2\vec{G}^{R}_0(\varepsilon,\textbf{p})\left(G^{R0}_0(\varepsilon,\textbf{p})\Sigma^{R0}_{B}(\varepsilon)+
  \vec{G}^{R}_0(\varepsilon,\textbf{p})\cdot\vec{\Sigma}^{R}_{B}(\varepsilon)\right),
  \label{eq:two-band:G^R,Ai:E}
  \end{eqnarray}
  \label{eq:two-band:G^R,A:E}
\end{subequations}
and the inner product of two vectors are defined as
\begin{eqnarray}
 \vec{A}\cdot\vec{B}&=&\sum_{\l=x,y,z}A^lB^l.
  \label{eq:Rashba:Born_all:Sigma^R,A:E2}
\end{eqnarray}

From Eqs.~(\ref{eq:Sigma^R,A:b}) and ~(\ref{eq:g^R,A:0}), we write the
self-energy $\hat{\Sigma}^{R}_{B}(\varepsilon)$ as
\begin{eqnarray}
 \hat{\Sigma}^R_{B}(\varepsilon) &=&  \Sigma^{R0}_{B}(\varepsilon )\hat{\sigma}^{0}+ \sum_{\l=x,y,z} \Sigma^{Rl}_{B}(\varepsilon )\hat{\sigma}^l ,
  \label{eq:Rashba:Born_all:Sigma^R,A:E}
\end{eqnarray}
with
\begin{subequations}
\begin{eqnarray}
  \Sigma^{R0}_{B}(\varepsilon) &=& n_{\text{imp}}u_{\text{imp}}^2\left((1-u_{\text{imp}}g^{R0}_0(\varepsilon))^2-u_{\text{imp}}^2g^{Rz}_0(\varepsilon)^2\right)^{-2}
  \nonumber\\
  &&\times\left[\left((1-u_{\text{imp}}g^{R0}_0(\varepsilon))^2+u_{\text{imp}}^2g^{Rz}_0(\varepsilon)^2\right)g^{R0}_{B}(\varepsilon)
    +2(1-u_{\text{imp}}g^{R0}_0(\varepsilon))u_{\text{imp}}g^{Rz}_0(\varepsilon)g^{Rz}_{B}(\varepsilon)\right],
  \label{eq:Rashba:Born_all:Sigma^R,A:E:0}\\
  \Sigma^{Rz}_{B}(\varepsilon) &=& n_{\text{imp}}u_{\text{imp}}^2\left((1-u_{\text{imp}}g^{R0}_0(\varepsilon))^2-u_{\text{imp}}^2g^{Rz}_0(\varepsilon)^2\right)^{-2}
  \nonumber\\
  &&\times\left[\left((1-u_{\text{imp}}g^{R0}_0(\varepsilon))^2+u_{\text{imp}}^2g^{Rz}_0(\varepsilon)^2\right)g^{Rz}_{B}(\varepsilon)
     +2(1-u_{\text{imp}}g^{R0}_0(\varepsilon))u_{\text{imp}}g^{Rz}_0(\varepsilon)g^{R0}_{B}(\varepsilon)\right],
  \label{eq:Rashba:Born_all:Sigma^R,A:E:z}\\
  \Sigma^{Ri}_{B}(\varepsilon) &=&
  n_{\text{imp}}u_{\text{imp}}^2\left((1-u_{\text{imp}}g^{R0}_0(\varepsilon))^2-
  u_{\text{imp}}^2g^{Rz}_0(\varepsilon)^2\right)^{-1}g^{Ri}_{B}(\varepsilon),
  \label{eq:Rashba:Born_all:Sigma^R,A:E:i}
\end{eqnarray}
\label{eq:Rashba:Born_all:Sigma^R,A:E}
\end{subequations}
and we have
\begin{eqnarray}
  g^{R\alpha}_{0,B}(\varepsilon)&=&\int\frac{d^2\textbf{p}}{(2\pi\hbar)^2} G^{R\alpha}_{0,B},
  \label{eq:Rashba:Sigma^R,A0,z:E}
\end{eqnarray}
where $\alpha\in\{0,x,y,z\}$.
The zeroth order components $G^{R\alpha}_0$ are computed as in appendix~\ref{app:sigma^R-0} and are used as input for the above
equations.

\section{The particle density}
\label{edensity}

Here, we present the derivation of Eq.(\ref{density}). In the absence of disorder scattering,
\begin{equation}
  \hat{G}^{R,A}_0(\varepsilon,\textbf{p})=[\varepsilon-\hat{H}_0(\textbf{p})\pm i0^+]^{-1}.
  \label{eq:G^R,A:free}
\end{equation}
At zero temperature, plugging Eq.~(\ref{eq:G^R,A:free}) into Eqs.~(\ref{eq:G^<:b,II}) and~(\ref{eq:number}), we can obtain
\begin{eqnarray}
  &&\bm n_e
  =-\int\!\frac{d\varepsilon}{\pi}\int\!\frac{d^2\textbf{p}}{(2\pi\hbar)^2}\left\{\sum_n\frac{1}{\epsilon-\epsilon_{n\textbf{p}}+i0^+}
 +e\hbar B\sum_{nm}\frac{1}{(\epsilon-\epsilon_{n\textbf{p}}+i0^+)^2}\frac{1}{\epsilon-\epsilon_{m\textbf{p}}+i0^+}
   \right.\nonumber\\
  &&\times \left.\frac{}{}\Im[\langle u_{n\bm \textbf{p}}|\hat{v}_x(\textbf{p})|u_{m\bm
  \textbf{p}}\rangle\langle u_{m\bm \textbf{p}}|\hat{v}_y(\textbf{p})|u_{n\textbf{p}}\rangle]\right\}.
\label{eq:eddetail}
\end{eqnarray}
$u_{n\textbf{p}}$ are the eigenfunctions of the unperturbed Hamiltonian
and $\epsilon_{n\textbf{p}}$ the eigenvalues. The integral over
$\epsilon$ contains simple and double poles. Using the residue theorem~\cite{Nunner}, we obtain
\begin{eqnarray}
  &&\bm n_e
  =\int\!\frac{d^2\textbf{p}}{(2\pi\hbar)^2}\sum_{n,occ} \{1
  +2ie\hbar B\sum_m
\Im[\langle u_{n\bm \textbf{p}}|\hat{v}_x(\textbf{p})|u_{m\bm
  \textbf{p}}\rangle\langle u_{m\bm \textbf{p}}|\hat{v}_y(\textbf{p})|u_{n\textbf{p}}\rangle]\},
\label{eq:eddetai2}
\end{eqnarray}
where $occ$ denotes summing over occupied states. Further simplification can be made by using the Sternheimer equation
\begin{eqnarray}
\hat{\upsilon}_j (\textbf{p})|u_{n\textbf{p}}\rangle
=(\epsilon_{n\textbf{p}}-\epsilon_{n^{\prime}\textbf{p}})|\frac{\partial
u_{n\textbf{p}} }{ \partial p_j }\rangle +\frac{\partial
\epsilon_{n\textbf{p}}}{\partial p_j}|u_{n\textbf{p}}\rangle,
\label{eq:stern}
\end{eqnarray}
and we finally arrive at the equation
\begin{eqnarray}
\bm n_e &=& \sum_{n,occ}\int\!\frac{d^2\textbf{p}}{(2\pi\hbar)^2}\left[ 1 + \frac{e}{\hbar}\textbf{B}\cdot\bm\Omega_{n}(\textbf{p})\right]. \label{eq:edapp}
\end{eqnarray}

\section{Orbital magnetization in the clean limit}
\label{integralom}

The derivations of Eq.(\ref{cleanM}) for the OM in the clean limit are present below.
When the relaxation rate vanishes, substituting Eq.~(\ref{eq:G^R,A:free}) into Eq.~(\ref{eq:IOM:int}),
we can write Eq.~(\ref{eq:IOM:int}) as
\begin{eqnarray}
  &&\bm M
  =e\hbar \int\!\frac{d\varepsilon}{2\pi}f(\varepsilon)\int\!\frac{d^2\textbf{p}}{(2\pi\hbar)^2}
\sum_{nm}(\epsilon_{n\bm \textbf{p}}-\mu)\Im[\langle u_{n\bm \textbf{p}}|\hat{v}_x(\textbf{p})|u_{m\bm
  \textbf{p}}\rangle\langle u_{m\bm \textbf{p}}|\hat{v}_y(\textbf{p})|u_{n\textbf{p}}\rangle]
   \nonumber\\
&&\times\left[\frac{1}{(\epsilon-\epsilon_{n\textbf{p}}+i0^+)^2}
\frac{1}{\epsilon-\epsilon_{m\textbf{p}}+i0^+}
-\frac{1}{(\epsilon-\epsilon_{n\textbf{p}}+i0^+)^2}
\frac{1}{\epsilon-\epsilon_{m\textbf{p}}+i0^+}\right] \label{eq:omfree}
\end{eqnarray}
Using the residue theorem, we find that
\begin{eqnarray}
  &&\bm M
  =-e\hbar\int\!\frac{d^2\textbf{p}}{(2\pi\hbar)^2}\times[\frac{f(\epsilon_{m \textbf{p}})-f(\epsilon_{n \textbf{p}})}{(\epsilon_{m\textbf{p}}-\epsilon_{n\textbf{p}})^2}
 +\frac{f^{\prime}(\epsilon_{n\textbf{p}})}{\epsilon_{n\textbf{p}}-\epsilon_{m\textbf{p}}}]
 \times\Im\sum_{nm}(\epsilon_{n\textbf{p}}-\mu)[\langle u_{n\textbf{p}}|\hat{v}_x(\textbf{p})|u_{m\bm
  \textbf{p}}\rangle\langle u_{m\textbf{p}}|\hat{v}_y(\textbf{p})|u_{n\textbf{p}}\rangle]
\label{eq:omfree2}
\end{eqnarray}
where $f^{\prime}_{n\textbf{p}}\equiv\partial f(\epsilon_{n\textbf{p}})/\partial\epsilon_{n\textbf{p}}$. With the help of the Sternheimer equation Eq.~(\ref{eq:stern}), we obtain
\begin{eqnarray}
  &&\bm M =\frac{i}{2}e\hbar\int\!\frac{d^2\textbf{p}}{(2\pi\hbar)^2}\sum_{n}
 \left[(\epsilon_{n\bm \textbf{p}}-\mu)\langle{\frac{\partial u_{n\textbf{p}}}
      {\partial\textbf{p}}|[\epsilon_{n\textbf{p}}-\hat{H}_{0}(\textbf{p})]
      \times|\frac{\partial u_{n\textbf{p}}}{\textbf{p}}}\rangle f_{n\bm
      p}^{\prime}
-\langle{\frac{\partial u_{n\textbf{p}}}{\partial \textbf{p}}|
      [\epsilon_{n\textbf{p}}+\hat{H}_{0}(\textbf{p})-2\mu]
      \times|\frac{\partial u_{n\textbf{p}}}{\partial \textbf{p}}}\rangle
    f_{n\textbf{p}}\right]\big|_z.
      \nonumber\\\label{eq:MK-finall}
  \end{eqnarray}
The above result can be written as
\begin{eqnarray}
\bm M &=& \sum_{n\textbf{p}}\left\{ \bm m_{n}(\textbf{p})f_{n\textbf{p}}+(\epsilon_{n\textbf{p}}-\mu)\bm m_{n}(\textbf{p})f_{n\textbf{p}}^{\prime}
 - \frac{e}{\hbar}(\epsilon_{n\textbf{p}}-\mu)\bm\Omega_{n \textbf{p}}(\textbf{p})\right\},
   \label{eq:MT-final}
\end{eqnarray}
where
$\bm m_{n}(\textbf{p})\!=\!(e/2\hbar)i\langle\bm\nabla_{\textbf{p}}u_{n\textbf{p}}|
[\epsilon_{n}(\textbf{p})-\hat{H}_{0}(\textbf{p})]\times|\bm\nabla_{\textbf{p}}u_{n\textbf{p}}\rangle$ is the orbital moment
of state $n, \textbf{p}$ and $\bm\Omega_{n}(\textbf{p}) \!=\!
i\langle{\bm\nabla_{\textbf{p}}u_{n\textbf{p}}|\times|\bm\nabla_{\textbf{p}}u_{n\textbf{p}}}\rangle$ is the Berry curvature. At zero temperature, $f^\prime$ becomes a $\delta$-function of $(\epsilon_{n\textbf{p}}-\mu)$, therefore we have in this case
\begin{eqnarray}
\bm M &=& \sum_{n\textbf{p}}\left[ \bm m_{n}(\textbf{p})f_{n\textbf{p}}
 - \frac{e}{\hbar}(\epsilon_{n\textbf{p}}-\mu)\bm\Omega_{n}(\textbf{p})\right]. \label{eq:MT-final2app}
\end{eqnarray}

\end{widetext}

\end{document}